\documentclass[aps,twocolumn,groupedaddress,showpacs]{revtex4}%
\usepackage{amsmath}
\usepackage{graphicx}%
\usepackage{amsfonts}%
\usepackage{amssymb}

\begin{document}

\title{The origin of noncommutativity?}
\author{Wenli He}
\affiliation{Institute of Modern Physics, Northwest University, Xian 710069, China}
\author{Liu Zhao}\thanks{Correspondence author}
\affiliation{Institute of Modern Physics, Northwest University, Xian 710069, China}
\date{November 5, 2001 }

\vspace{1cm}

\begin{abstract}
Consistent boundary Poisson structures for open string theory coupled to
background $B$-field are considered using the new approach proposed in
hep-th/0111005. It is found that there are infinitely many consistent Poisson
structures, each leads to a consistent canonical quantization of open string
in the presence of background $B$-field. Consequently, whether the $D$-branes
to which the open string end points are attached is noncommutative or not
depends on the choice of a particular Poisson structure.
\end{abstract}
\maketitle


%Optional argument for running titles on pages
%\title[]{}

%repeat the \author .. \affiliation  etc. as needed

%\affiliation command applies to all authors since the last

%\hspace{1cm}
%\email{lzhao@nwu.edu.cn}

%\email{hewenli@phy.nwu.edu.cn}

%\date{October 15,2001}

%insert suggested PACS numbers in braces on next line

%insert suggested keywords - APS authors don't need to do this


These days it is widely believed that quantization of open string theory in
the presence of background NS $B$-field would give rise to noncommutativity
at the boundary, known as $D$-branes. This observation has led to considerable
interests in the study of noncommutative field theories living on $D$-branes,
see e.g. \cite{NK} for review. Among various different approaches in deriving the
noncommutativity, there are the Seiberg-Witten approach \cite{SW} (who
extracted the commutation relations between spacetime coordinates from the
open string propagator \cite{Callan} using the idea given in \cite{Schomerus}%
), the Dirac approach with \cite{ModDirac1}\cite{ModDirac}\cite{ModDirac2}\cite{Ion} or
without \cite{NDirac} modifications, symplectic quantization \cite{symp}
and so on, and there are some mysterious
discrepancies between the results obtained from different approaches.

\bigskip

In this paper, we shall use yet another approach proposed recently by us in
\cite{ZH} to quantize open string in the background $B$-field. This approach
is based on a consistent definition of canonical Poisson structure for field
theories with boundaries, according to the causality and locality analysis. We
find that, following our approach, the apparent discrepancies arisen from
different approaches acquire a natural explanation - the discrepancies are
just a reflection of the fact that there are many (and actually infinitely
many) consistent Poisson structures for the world sheet theory of open string
in the presence of background $B$-field. In this view point, none of the
results for spacetime noncommutativity obtained thus far should be considered
superior to others, and whether field theories living on $D$-branes is
commutative or noncommutative depends on which Poisson structure the observer chooses.

\bigskip

Now let us formulate our approach in detail. The world sheet action of open
string theory coupled to a constant background NS $B$-field and background
$U(1)$ gauge fields $A$ reads%
\begin{align}
S  &  =\frac{1}{4\pi\alpha^{\prime}}\nonumber\\
&  \times\int d^{2}\sigma\left[  g^{ab}\eta_{ij}\partial_{a}X^{i}\partial
_{b}X^{j}+2\pi\alpha^{\prime}B_{ij}\epsilon^{ab}\partial_{a}X^{i}\partial
_{b}X^{j}\right] \nonumber\\
&  +\left.  \int d\tau A_{i}\partial_{\tau}X^{i}\right|  _{\sigma=\pi}-\left.
\int d\tau A_{i}\partial_{\tau}X^{i}\right|  _{\sigma=0}, \label{action1}%
\end{align}
where $g^{ab}=diag(-1,1),\epsilon^{01}=1=-\epsilon^{10},B_{ij}=-B_{ji}$ and
$\eta=diag(-,+,...,+)$. In the case when both ends of the string are attached
to the same brane, the last two boundary terms can be rewritten as
$-\frac{1}{2\pi\alpha^{\prime}}\int d^{2}\sigma F_{ij}\epsilon^{ab}%
\partial_{a}X^{i}\partial_{b}X^{j}$ and the action (\ref{action1}) becomes%
\begin{align}
S  &  =\frac{1}{4\pi\alpha^{\prime}}\label{action2}\\
&  \times\int d^{2}\sigma\left[  g^{ab}\eta_{ij}\partial_{a}X^{i}\partial
_{b}X^{j}+2\pi\alpha^{\prime}\mathcal{F}_{ij}\epsilon^{ab}\partial_{a}%
X^{i}\partial_{b}X^{j}\right]  ,
\end{align}
with $\mathcal{F}=B-F=B-dA$, which is invariant under $U(1)$ gauge transform
as well as $\Lambda$ translation defined as $B\rightarrow B+d\Lambda
,A\rightarrow A+\Lambda$. For simplicity we set $F_{ij}=0$ throughout this article.

\bigskip

The variation of (\ref{action2}) yields the equation of motion%
\begin{equation}
(\partial_{\tau}^{2}-\partial_{\sigma}^{2})X^{i}=0 \label{eqm}%
\end{equation}
and the mixed boundary conditions%
\begin{equation}
\eta_{ij}\partial_{\sigma}X^{j}+2\pi\alpha^{\prime}B_{ij}\partial_{\tau}%
X^{j}|_{\sigma=0,\pi}=0. \label{boundary}%
\end{equation}

The canonical conjugate momenta are given as%
\[
P_{i}\equiv\frac{\delta L}{\delta\partial_{\tau}X^{i}}=
\frac{1}{2\pi\alpha'}\left(\partial_{\tau}%
X_{i}+2\pi\alpha^{\prime}B_{ij}\partial_{\sigma}X^{j}\right),
\]
which, under the naive canonical Poisson structure, obey%
\begin{align}
\{X^{i}(\sigma),X^{j}(\sigma^{\prime})\}  &  =\{P_{i}(\sigma),P_{j}%
(\sigma^{\prime})\}=0,\label{pbnaive1}\\
\{X^{i}(\sigma),P_{j}(\sigma^{\prime})\}  &  =\delta_{j}^{i}\delta
(\sigma-\sigma^{\prime}). \label{pbnaive2}%
\end{align}
However, due to the appearance of the boundary conditions (\ref{boundary}),
the naive Poisson structure (\ref{pbnaive1},\ref{pbnaive2}) does not hold
consistently, and the major task one has to fulfil in order to quantize the
system (\ref{eqm},\ref{boundary}) is to obtain a Poisson structure which is
consistent with (\ref{boundary}).

\bigskip

Now let us remind that the boundary conditions (\ref{boundary}) are
constraints only at the end points of the open string, they make the naive
Poisson structure inconsistent only at the end points also. According to the
principle of locality, to make the Poisson structure consistent with the
boundary conditions, one only needs to modify the naive Poisson brackets at
the string end points. Therefore, we postulate the following form for the
consistent Poisson structure,
\begin{align}
&  \{X^{i}(\sigma),X^{j}(\sigma^{\prime})\}\nonumber\\
&  =(\mathcal{A}_{L})^{ij}\delta(\sigma+\sigma^{\prime})+(\mathcal{A}%
_{R})^{ij}\delta(2\pi-\sigma-\sigma^{\prime}),\label{PB1}\\
&  \{X^{i}(\sigma),P_{j}(\sigma^{\prime})\}\nonumber\\
&  =\delta_{j}^{i}\delta(\sigma-\sigma^{\prime})+(\mathcal{B}_{L})_{j}%
^{i}\delta(\sigma+\sigma^{\prime})+(\mathcal{B}_{R})_{j}^{i}\delta(2\pi
-\sigma-\sigma^{\prime}),\label{PB2}\\
&  \{P_{i}(\sigma),P_{j}(\sigma^{\prime})\}\nonumber\\
&  =(\mathcal{C}_{L})_{ij}\delta(\sigma+\sigma^{\prime})+(\mathcal{C}%
_{R})_{ij}\delta(2\pi-\sigma-\sigma^{\prime}), \label{PB3}%
\end{align}
where for the moment $\mathcal{A}_{L,R}$, $\mathcal{B}_{L,R}$, $\mathcal{C}%
_{L,R}$ are still unknown, but they are assumed to be some operators which may
act on the variable $\sigma^{\prime}$ and are antisymmetric in
$i\leftrightarrow j$. The delta functions $\delta(\sigma+\sigma^{\prime})$ and
$\delta(2\pi-\sigma-\sigma^{\prime})$ are non-vanishing only at the string end
points. The first delta function term in (\ref{PB2}) has to be there in order
that the equation of motion (\ref{eqm}) follow from the canonical formalism.

\bigskip

To determine the values of the operators $\mathcal{A}_{L,R}$, $\mathcal{B}%
_{L,R}$, $\mathcal{C}_{L,R}$, we now reformulate the boundary conditions
(\ref{boundary}) as the following constraints,%
\begin{align*}
&  (G_{L})^{i}\\
&  =\int d\sigma\delta(\sigma)\left[  (2\pi\alpha^{\prime})^{2}B^{ij}%
(P_{j}-B_{jk}\partial_{\sigma}X^{k})+\eta^{ij}\partial_{\sigma}X_{j}\right] \\
&  \simeq0,\\
&  (G_{R})^{i}\\
&  =\int d\sigma\delta(\pi-\sigma)\left[  (2\pi\alpha^{\prime})^{2}%
B^{ij}(P_{j}-B_{jk}\partial_{\sigma}X^{k})+\eta^{ij}\partial_{\sigma}%
X_{j}\right] \\
&  \simeq0,
\end{align*}
where we adopt the convention (for explanation, see \cite{ZH})%
\[
\int_{0}^{\pi}d\sigma\delta(\sigma)=\int_{0}^{\pi}d\sigma\delta(\pi
-\sigma)=1.
\]
Notice that the boundary constraints $G_{L,R}$ contain both $X^{i}$ and
$P_{j}$, that is why all three of the naive Poisson brackets are supposed to
be modified, in contrast to the cases studied in \cite{ZH}, where only the
$\{\varphi,\pi\}$ bracket gets modified.

\bigskip

Now straightforward calculations using (\ref{PB1}-\ref{PB3}) show that the
following Poisson brackets hold,%
\begin{align}
&  \{(G_{L})^{i},X^{j}(\sigma^{\prime})\}\nonumber\\
\qquad &  =\left[  -(2\pi\alpha^{\prime})^{2}B(I+\mathcal{B}_{L}%
+B\mathcal{A}_{L}\partial_{\sigma^{\prime}})+\mathcal{A}_{L}\partial
_{\sigma^{\prime}}\right]  ^{ij}\delta(\sigma^{\prime}),\label{consist1}\\
&  \{(G_{L})^{i},P_{j}(\sigma^{\prime})\}\nonumber\\
&  =[(2\pi\alpha^{\prime})^{2}B(\mathcal{C}_{L}-B(-I+\mathcal{B}_{L}%
)\partial_{\sigma^{\prime}})\nonumber\\
&  +(-I+\mathcal{B}_{L})\partial_{\sigma^{\prime}}]_j^i\delta(\sigma^{\prime
}).\label{consist2}%
\end{align}
Similar Poisson brackets for $G_{R}$ also hold, with only the replacement
$\delta(\sigma^{\prime})\rightarrow\delta(\pi-\sigma^{\prime})$. While doing
the above calculations, we have assumed that $\mathcal{A}_{L,R}$,
$\mathcal{B}_{L,R}$, $\mathcal{C}_{L,R}$ do not explicitly depend on
$\sigma^{\prime}$ (hence commute with $\partial_{\sigma^{\prime}}$). In order
that the Poisson brackets (\ref{PB1}-\ref{PB3}) be consistent with the
boundary conditions, the Poisson brackets (\ref{consist1},\ref{consist2}) and
their counterparts for $G_{R}$ have to vanish. Since the delta functions
$\delta(\sigma^{\prime})$ and $\delta(\pi-\sigma^{\prime})$ do not vanish
identically along the open string, we have to set the operators acting on them
to be zero. This leads to the following operator equations which are helpful
in determining the operators $\mathcal{A}_{L,R}$, $\mathcal{B}_{L,R}$,
$\mathcal{C}_{L,R}$,%
\begin{align}
&  (2\pi\alpha^{\prime})^{2}B(I+\mathcal{B}_{L,R}+B\mathcal{A}_{L,R}%
\partial_{\sigma^{\prime}})-\mathcal{A}_{L,R}\partial_{\sigma^{\prime}%
}=0,\label{sys1}\\
&  (2\pi\alpha^{\prime})^{2}B[\mathcal{C}_{L,R}-B(-I+\mathcal{B}%
_{L,R})\partial_{\sigma^{\prime}}]\nonumber\\
&  \qquad+(-I+\mathcal{B}_{L,R})\partial_{\sigma^{\prime}}=0.\label{sys2}%
\end{align}
This is a system of two equations for three unknowns, meaning that the
consistent values of the operators $\mathcal{A}_{L,R}$, $\mathcal{B}_{L,R}$,
$\mathcal{C}_{L,R}$ are not unique. One may be tempted to use the conditions%
\begin{equation}
\{(G_{L,R})^{i},(G_{L,R})^{j}\}=0\label{Gij}%
\end{equation}
to derive further restrictions over the operators $\mathcal{A}_{L,R}$,
$\mathcal{B}_{L,R}$, $\mathcal{C}_{L,R}$. However this is impossible, because
the conditions $\{(G_{L,R})^{i},X^{j}(\sigma^{\prime})\}=0,\{(G_{L,R}%
)^{i},P_{j}(\sigma^{\prime})\}=0$ automatically ensure the condition
(\ref{Gij}) according to Jacobi identity. In fact, since the set of variables
$X^{i},P_{j}$ constitute a complete set of independent degrees of freedom in
the phase space, the vanishing of the Poisson brackets $\{(G_{L,R})^{i}%
,X^{j}(\sigma^{\prime})\},$ $\{(G_{L,R})^{i},P_{j}(\sigma^{\prime})\}$
automatically guarantees the vanishing of Poisson brackets between $G_{L,R}$
and everything in the phase space. Therefore, there are no further
restrictions to the operators $\mathcal{A}_{L,R}$, $\mathcal{B}_{L,R}$,
$\mathcal{C}_{L,R}$. The canonical quantization of open string coupled to
background $B$-field is then accomplished by replacing the Poisson brackets
(\ref{PB1}-\ref{PB3}) by the corresponding commutation relations.

\bigskip

Following our discussions made above, we may conclude that \emph{there are
infinitely many consistent Poisson structures for open string theory coupled
to background }$B$\emph{-field, each leads to a consistent canonical
quantization of the theory. }Whether one sees noncommutativity on the branes
to which the open string is attached depends on the choice of particular
solutions to the system (\ref{sys1},\ref{sys2}). Given any one of the
operators $\mathcal{A}_{L,R}$, $\mathcal{B}_{L,R}$, $\mathcal{C}_{L,R}$, one
may get a particular solution for the other two. This may explain the apparent
discrepancies on the quantization of open string under background $B $-field.

\bigskip

To be more specific, let us now demonstrate some particular solutions to the
system (\ref{sys1},\ref{sys2}). The most simple solutions may come about when
one of the three unknowns is set to zero. There are three such solutions:

\begin{enumerate}
\item $\mathcal{A}_{L,R}=0$, which corresponds to \emph{commutative} branes.
In this case, we have $\mathcal{B}_{L,R}=-I$, $\mathcal{C}_{L,R}%
=\frac{2}{(2\pi\alpha^{\prime})^{2}}B^{-1}[I-(2\pi\alpha^{\prime})^{2}B^{2}]
\partial_{\sigma^{\prime}}$, or in component form%
\begin{align*}
(\mathcal{B}_{L,R})_{j}^{i}  &  =-\delta_{j}^{i},\\
(\mathcal{C}_{L,R})_{ij}  &  =\frac{2}{(2\pi
\alpha^{\prime})^{2}}\left[  (\eta+2\pi\alpha^{\prime}B)\frac{1}{B}(\eta
-2\pi\alpha^{\prime}B)\right]  _{ij}\partial_{\sigma^{\prime}};
\end{align*}

\item $\mathcal{B}_{L,R}=0$. This leads to the solution $\mathcal{A}%
_{L,R}=(2\pi\alpha^{\prime})^{2} B [I-(2\pi\alpha^{\prime})^{2}B^{2}]^{-1}
\partial_{\sigma^{\prime}}^{-1},$ $\mathcal{C}_{L,R}=\frac{1}{(2\pi\alpha^{\prime
})^{2}} [I-(2\pi\alpha^{\prime})^{2}B^{2}] {B^{-1}}\partial
_{\sigma^{\prime}}$, or%
\begin{align*}
(\mathcal{A}_{L,R})^{ij}  &  =(2\pi\alpha^{\prime})^{2}
\left[  \frac{1}{\eta+2\pi\alpha^{\prime}B}B\frac{1}%
{\eta-2\pi\alpha^{\prime}B}\right]  ^{ij}\partial
_{\sigma^{\prime}}^{-1},\\
(\mathcal{C}_{L,R})_{ij}  &  =\frac{1}{(2\pi
\alpha^{\prime})^{2}}\left[  (\eta+2\pi\alpha^{\prime}B)\frac{1}{B}(\eta
-2\pi\alpha^{\prime}B)\right]  _{ij}\partial_{\sigma^{\prime}};
\end{align*}

\item $\mathcal{C}_{L,R}=0$. Then $\mathcal{A}_{L,R}=2(2\pi
\alpha^{\prime})^{2}[I-(2\pi\alpha^{\prime})^{2}B^{2}]^{-1} B
\partial_{\sigma^{\prime}}^{-1},$ $\mathcal{B}_{L,R}=I$, or%
\begin{align*}
(\mathcal{A}_{L,R})^{ij}  &  =2(2\pi\alpha^{\prime})^{2}
\left[  \frac{1}{\eta+2\pi\alpha^{\prime}B}B\frac{1}%
{\eta-2\pi\alpha^{\prime}B}\right]  ^{ij}\partial
_{\sigma^{\prime}}^{-1}
,\\
(\mathcal{B}_{L,R})_{j}^{i}  &  =\delta_{j}^{i}.
\end{align*}
\end{enumerate}

In contrast to the previous results
\cite{ModDirac1}\cite{ModDirac}\cite{ModDirac2}\cite{Ion}\cite{NDirac} (which make use
of the discretised or Fourier transformed version of the coordinate fields fields and
their conjugate momenta), our results (using the unmodified coordinate fields and standard
conjugate momenta) for the Poisson brackets
$\{X^i(\sigma),X^j(\sigma')\}$ etc contain explicitly the factor
$\partial_{\sigma'}^{-1}\delta(\sigma+\sigma')$ (resp.
$\partial_{\sigma'}^{-1}\delta(2\pi-\sigma-\sigma')$) or
$\partial_{\sigma'}\delta(\sigma+\sigma')$ (resp.
$\partial_{\sigma'}\delta(2\pi-\sigma-\sigma')$) at the string end points (the
operator $\partial_{\sigma'}^{-1}$ has well-defined meaning in the momentum space).
The use of these factors enables us to write the Poisson brackets at any point along
the open string in a consistent and unified way. Moreover, apart from these factors,
the results for noncommutativity given in case 2 agree with the results
of \cite{ModDirac}, i.e.
\begin{eqnarray*}
\{X^i(0),X^j(0)\} &\propto& (2\pi\alpha^{\prime})^{2}
\left[  \frac{1}{\eta+2\pi\alpha^{\prime}B}B\frac{1}%
{\eta-2\pi\alpha^{\prime}B}\right]  ^{ij},\\
\{P^i(0),P^j(0)\} &\propto& \frac{1}{(2\pi
\alpha^{\prime})^{2}}\left[  (\eta+2\pi\alpha^{\prime}B)\frac{1}{B}(\eta
-2\pi\alpha^{\prime}B)\right]  _{ij}
\end{eqnarray*}
(cf. equations (22-24, 30) of \cite{ModDirac}), while the result of case 3 is
basically the same as the result of \cite{ModDirac2},
\begin{eqnarray*}
\{X^i(0),X^j(0)\} &\propto& 2(2\pi\alpha^{\prime})^{2}
\left[  \frac{1}{\eta+2\pi\alpha^{\prime}B}B\frac{1}%
{\eta-2\pi\alpha^{\prime}B}\right]  ^{ij},\\
\{P^i(0),P^j(0)\} &=& 0
\end{eqnarray*}

\noindent (notice that the normalization for the background field $B$ in \cite{ModDirac2}
differs from ours by a factor of $2\pi\alpha'$).

\bigskip

Of course in the above analysis, one need not take the same solution for
the left and right end points of the open string. Rather, one may take
different particular solutions for the two boundaries, and, in particular, one
may think about the case in which one end of the open string is attached to a
noncommutative brane while the other is attached to a commutative one. This
may lead to the interesting question about the interaction between a
noncommutative brane and an ordinary one. In a word, the quantization of open
string theory coupled with background NS $B$-field contains far richer contents
than expected. More careful study on this problem is still needed in order to
have full understanding of this problem.

This work is supported by the National Natural Science Foundation of china.

\vspace{0.3cm}

%\bibliography{}

\end{document}